\documentclass[12pt]{article} 
\usepackage{epsfig}
\usepackage{amsmath,graphicx}
\usepackage{cite}
\usepackage{color}

\newcommand{\be}{\begin{equation}}
\newcommand{\ee}{\end{equation}}
\newcommand{\bea}{\begin{eqnarray}}
\newcommand{\eea}{\end{eqnarray}}
\newcommand{\beq}{\begin{equation}}
\newcommand{\eeq}{\end{equation}}
\newcommand{\ba}{\begin{array}}
\newcommand{\ea}{\end{array}}
\newcommand{\beqa}{\begin{eqnarray}}
\newcommand{\eeqa}{\end{eqnarray}}
\newcommand{\lsim}{\stackrel{<}{_\sim}}
\newcommand{\gsim}{\stackrel{>}{_\sim}}
\newcommand{\cA}{{\cal A}}
\newcommand{\cO}{{\mathcal{O}}}
\newcommand{\cB}{{\cal B}}

\newcommand{\dis}{\displaystyle}
\newcommand{\no}{\nonumber}

\def\gappeq{\mathrel{\rlap {\raise.5ex\hbox{$>$}}
{\lower.5ex\hbox{$\sim$}}}}
\def\lappeq{\mathrel{\rlap{\raise.5ex\hbox{$<$}}
{\lower.5ex\hbox{$\sim$}}}}

\def\slash{ \! \! \! \!  /~}

\def\bea{\begin{eqnarray}}   
\def\eea{\end{eqnarray}}

\def\tev{{\rm TeV}}

\begin{document}
%%%%%%%%%%%%%%%%%%%%%%%%%%%%%%%%%%%%%%%%%%%%%%%%%%%%%%%%%%%%%%%%%%%%%%%

\thispagestyle{empty}
\begin{flushright}
TUM-HEP-677/07\\
20 November 2007
\end{flushright}
\vskip 1.5 true cm 

\begin{center}
{\Large\bf  
  Solving the flavour problem with \\ [5pt]
  hierarchical fermion wave functions}
 \\ [25 pt]
{\sc  Sacha Davidson${}^{a}$, Gino Isidori${}^{b}$, Selma Uhlig${}^{c}$}
 \\ [25 pt]
{\sl ${}^a$ IPN de Lyon, Universit\'e  Lyon 1, CNRS,
            F-69622 Villeurbanne, CEDEX, France } \\ [5 pt] 
{\sl ${}^b$ Scuola Normale Superiore, Piazza dei Cavalieri 7 I-56100 Pisa, Italy\\
           and INFN, Laboratori Nazionali di Frascati, Via E. Fermi 40, 
           I-00044 Frascati, Italy} \\ [5 pt] 
{\sl ${}^c$Technische Universit{\"a}t M{\"u}nchen, Physik Department,
           D-85747 Garching, Germany} \\ [20 pt] 
{\bf Abstract  } \\
\end{center} 
We investigate the flavour structure of generic extensions of the SM 
where quark and lepton mass hierarchies and the suppression 
of flavour-changing transitions originate only by the normalization 
constants of the fermion kinetic terms. We show that in such scenarios 
the contributions to quark FCNC transitions from dimension-six 
effective operators are sufficiently suppressed without 
(or with modest) fine tuning 
in the effective scale of new physics. The most serious 
challenge to this type of scenarios appears in the lepton 
sector, thanks to the stringent bounds on LFV. The phenomenological 
consequences of this scenarios in view of improved experimental 
data on quark and lepton FCNC transitions, and its differences 
with respect to the Minimal Flavour Violation hypothesis are also discussed.

\vskip 1.0  cm

\section{Introduction}

The Standard Model (SM) can be regarded as the low-energy limit of a 
general effective Lagrangian.
Within the renormalizable part of such a Lagrangian 
there are two types of operators involving fermion 
fields: kinetic and Yukawa terms.  
In general both these terms can have a non-trivial flavour structure. 
Since the description of physics is invariant under field reparametrizations,
it is only the relative flavour structure between 
kinetic and Yukawa terms that has a physical meaning.  

Employing the canonical normalization of the kinetic terms, 
the physical flavour structure of the SM (or the renormalizable 
part of the effective theory) is manifest in the Yukawa matrices.
As it is well known, these have a quite peculiar form: their eigenvalues are very 
hierarchical and the two matrices  in the quark
sector  are quasi-aligned in flavour space,
with the misalignment parameterized by the CKM matrix. 
This structure is responsible for the great successes of the SM
in the flavour sector, including the strong suppression 
of CP-violating and flavour-changing neutral current (FCNC)
amplitudes. 

Several additional flavour structures could appear 
in the tower of higher-dimensional operators which belongs to
the non-renormalizable part of the effective Lagrangian.
However, if we assume an effective scale of new physics 
in the TeV range (as expected by a  natural stabilization 
of the electro-weak symmetry-breaking sector), 
present experimental results 
leave a very limited room for new flavour structures.
A natural way out to this problem, the so-called flavour problem, is provided 
by the  hypothesis of minimal flavour violation 
(MFV)~\cite{MFV0,D'Ambrosio:2002ex,Cirigliano:2005ck}.
According to this hypothesis, 
there exists a flavour symmetry defined by canonically normalized kinetic terms, and the Yukawa matrices 
are its only breaking sources. 
In other words, the SM Yukawa matrices 
are treated as the only non-negligible spurions 
of the $SU(3)^5$ flavour symmetry~\cite{D'Ambrosio:2002ex}. 

The MFV hypothesis is a  simple  and effective solution 
to the flavour problem~\cite{Buras:2000dm,Bona:2007vi},
but is far from being the only  allowed possibility. 
Various alternatives or variations of this hypothesis
have indeed been proposed in the recent 
literature~\cite{Agashe:2005hk,Davidson:2006bd,Grinstein:2006cg,Feldmann:2006jk,BarShalom:2007pw}.
%%%\query{references modified}
The question we would like to address in this work is 
the viability, in general terms, of solutions to the flavour 
problem  based not on a flavour symmetry in the low-energy
effective theory,  but  instead  on hierarchies in
the kinetic terms which  suppress flavour-changing transitions.
This could be a  ``democratic'' alternative,  
where the many coupling constant matrices can be of any form,  
and the only restriction is that  the kinetic terms 
should  have a hierarchical normalization.
The wave function normalization factors  then  function as a
distorting lense, through which all interactions are seen
as approximately aligned on the magnification axes 
of the lense.

From a { model-building } point of view,
hierarchical fermion wave functions
(and corresponding hierarchical kinetic terms) can emerge 
in scenarios with extra dimensions, where the hierarchy in the 
four-dimensional wave functions reflect their non-trivial 
profile in the extra dimension~\cite{ArkaniHamed:1999dc,Gherghetta}.
In particular, a suppression of this type can be  
present in Randall-Sundrum (RS) scenarios~\cite{Randall:1999ee},
whose flavour phenomenology 
was studied in Ref.~\cite{Kitano:2000wr,Huberetal,preNMFV}.  
In the recent literature, variations of such models  have been
constructed with  particular  attention to the flavour 
structure~\cite{Contino:2006nn,Cacciapaglia:2007fw,Fitzpatrick:2007sa}.\footnote{~The 
scenario of Ref.~\cite{Contino:2006nn}, where the suppression 
of flavour-changing transitions arises only by the mixing 
of the SM fermions with the composite fields 
(in the dual four-dimensional description of a 
five-dimensional warped geometry) 
is a clear example of the class of models we intend 
to study. Although the scenario we  consider
emerges naturally in the RS framework, 
other  solutions to the flavour problem  have been proposed 
in the RS context. For instance,
additional symmetries can be imposed in warped geometry in order to recover 
a MFV structure in four~\cite{Cacciapaglia:2007fw}
or five dimensions~\cite{Fitzpatrick:2007sa}.}
{ Hierarchical fermion wave functions  can
also arise due to  Renormalisation Group running,
with large, positive, and  distinct anomalous dimensions
for the standard model fields of different generations
\cite{Nelson:2000sn}.}

It is always possible to redefine fields so as 
to choose a basis in which the Yukawa interactions 
are not hierarchical, and the SM flavour structure manifests
itself through the hierarchy of the kinetic terms.
As long as we look only at the renormalizable part of the
low-energy  effective Lagrangian, there is no way to isolate 
the origin of the flavour hierarchy (kinetic vs.
Yukawa terms). However, the different assumptions 
about its origin lead to different ans\"atze 
for the flavour structure of the higher-dimensional operators.  
In this paper we analyse the class of 
scenarios where dimensionless  coupling matrices (both Yukawas
and non-renormalizable operators) have $\cO (1)$
entries. That is, they 
do not exhibit a specific flavour structure 
in a  basis where the kinetic terms are 
hierarchical.   More explicitly,
we investigate whether it is possible to choose a hierarchy 
for the fermion wave functions such that, after 
moving to the canonical basis,
the contributions from dimension-six FCNC 
operators are sufficiently suppressed
{ and the Standard Model Yukawa hierachies are obtained}.
We deem the scenario to work if the 
dimensional suppression scale of the operators
turns out to be $\lsim 10$ TeV  (as in the MFV scenario).
{ This corresponds to one-loop contributions
from new particles of mass
$\sim 3 m_Z$ and SM gauge couplings.} 
We analyse this problem both in the quark and in the 
lepton sector, and we discuss the differences arising 
with respect to the MFV scenario.

The scenario we are considering has some similarities
with the Next-to-MFV framework of Ref.~\cite{Agashe:2005hk},
{ where flavoured couplings induced by New Physics(NP) 
are ``quasi-aligned'' to the SM Yukawas, and where
the NP couples ``dominantly'' to the third generation.
The hierarchical wave-function scenario is an 
example of how New Physics could interact dominantly 
with the third generation; however, this scenario 
differ from the NMFV hypothesis studied in 
Ref.~\cite{Agashe:2005hk}.}
There, the Lagrangian has a  $U(2)^3$
symmetry acting on the quarks of the first two generations
(broken only by the quark Yukawas) and new physics 
interactions involving the third generation are arbitrary.
The hierarchical wave functions we consider differ 
by not appealing to any flavour symmetry, or to a restricted  
set of spurions. This difference has { non-negligible} 
phenomenological implications in the case of rare transitions 
among the first two generations of quarks and leptons.
Yet another definition of ``dominant coupling to the
third generation'' is used  by the UTfit
Collaboration\cite{Bona:2007vi}, who take  $\Delta F = 2$ operators
of arbitrary chirality to all have
the same CKM-like suppression $\propto |V_{ti}^* V_{tj}|^2$. 
Our bounds differ also from those in 
Ref.~\cite{Bona:2007vi}, especially in the case 
of chirality-flippling operators, although in 
practice we find similar conclusions about 
the key role of $\epsilon_K$ in constraining 
both their and our scenario.

The paper is organized as follows: in Section~\ref{sect:basic}
we discuss the basic setup of our scenario 
for the quark sector. The corresponding bounds 
on the effective operators from 
quark FCNCs are analysed in  Section~\ref{sect:bounds}.
The extension to the lepton sector and the bounds 
from  lepton FCNC transitions are discussed
in Section~\ref{sect:leptons}. Section~\ref{sect:comparison}
is devoted to a general discussion of the bounds 
and a comparison with the MFV scenario.
The results are summarized in the Conclusions.

%%%%%%%%%%%%%%%%%%%%%%%%%%%%%%%%%%%%%%%%%%%%%%%%%%%%%%%%%%%%%%%%%%%%%%
\section{Basic setup for the quark sector}
\label{sect:basic}
%%%%%%%%%%%%%%%%%%%%%%%%%%%%%%%%%%%%%%%%%%%%%%%%%%%%%%%%%%%%%%%%%%%%%%

The operators involving quark fields 
in the renormalizable part of the effective Lagrangian are 
\beq
\label{lagragian}
{\cal L}_{\rm quarks}^{d=4} = 
\overline{Q_L} X_Q   D\slash Q_L + \overline{D_R}  X_D D\slash D_R + 
\overline{U_R} X_U D\slash U_R +
\overline{Q_L} Y_{D} D_R H +
\overline{Q_L} Y_{U} U_R H_c~,
\eeq
where  $H_c=i \tau_2 H^*$. In a generic non-canonical basis, 
the three $X_{Q,D,U}$ and the two $Y_{D,U}$ are $3\times 3$ complex matrices. 
The usual choice of  field normalization and
%A convenient reference 
basis is obtained by  diagonalising the 
kinetic terms, re-scaling the fields to obtain the 
canonical normalization of the 
kinetic terms, and finally  diagonalising  the down-quark masses.
With this choice 
\beq 
\label{eq:MFVbasis}
X_Q = X_D = X_U = I~, 
\qquad Y_{D}=\lambda_D~, \qquad Y_U=V^\dagger \lambda_U~,
\eeq
where $I$ is the identity matrix, $V$ is the CKM matrix and
\beq
\lambda_D=\frac{1}{v} {\rm diag}(m_d, m_s, m_b)~, 
\qquad \lambda_U=\frac{1}{v} {\rm diag} (m_u, m_c, m_t)~,
\eeq
with $v = (\langle H^\dagger H\rangle)^{1/2} = 174$ GeV.

{}For the purposes of this paper we express the
$d = 4$  Lagrangian with a different, hierarchical, field
normalization  for the kinetic terms, and then add the higher-dimensional
operators with ``democratic'' flavour couplings.
An example of the bases where the
Yukawas are no longer hierarchical can be reached 
starting from the basis (\ref{eq:MFVbasis}),  by performing
a unitary transformation  on $U_R$ and by an  appropriate  
rescaling of the fields.
In particular, denoting by $Z_A$~($A=Q,U,D$) the diagonal 
matrices by which we rescale the fermion fields
($Q_A \to Z_A^{-1} Q_A$)
and by $W_A$ the complex matrices describing their unitary transformations
in the canonical basis ($W_U^\dagger W_U=I$), we can move to a non-canonical 
basis where\footnote{~With different unitary transformation we could have 
chosen $Y_U \to I$ and $Y_D \to \cO(1)$, or $Y_{U,D} \to \cO(1)$.
As it will become clear in the next section, the choice in 
Eqs.~(\ref{eq:defZQD})--(\ref{eq:defZU}), is the simplest one 
for the phenomenological analysis of FCNC constraints.}
\bea 
Y_D  &\to&  Z_Q^{-1}\lambda_D Z_D^{-1}= I~, \qquad\qquad\quad 
   (I)_{ij} =\delta_{ij}~,  \label{eq:defZQD} \\
Y_U  &\to&  Z_Q^{-1} V^\dagger \lambda_U  W_U Z_U^{-1} = T_U \qquad 
   (T_U)_{ij}  = \cO(1)~. \label{eq:defZU}
\eea
In this new basis the hierarchical structure usually attributed to the 
Yukawa couplings is hidden in the flavour structure of the (diagonal) 
kinetic terms:
\beq
X_Q = Z_Q^{-2}~, \qquad  X_D = Z_D^{-2}~, \qquad  X_U =~ Z_U^{-2} ~. 
\label{eq:Xmaim}
\eeq
The conditions we have imposed on the $Z_A$ in 
Eqs.~(\ref{eq:defZQD}) and (\ref{eq:defZU}) 
do not specify completely their structure. 
However, assuming the maximal entries in the $Z_A$
are at most of $\cO(1)$ implies a hierarchical structure 
of the type
\beq
\label{parameterZ}
Z_A =  {\rm diag}(z_A^{(1)}, z_A^{(2)}, z_A^{(3)}  )~, \qquad 
z_A^{(1)}\ll z_A^{(2)}\ll z_A^{(3)} \lsim 1~.
\eeq

The framework we want to investigate is a scenario where
the hierarchical structure in Eq.~(\ref{parameterZ})
is the only responsible for the natural
suppression of FCNCs. More explicitly, we assume 
that with  hierarchical normalization of
 the kinetic terms, Eq.~(\ref{eq:Xmaim}), 
all the higher-dimensional operators of the effective Lagrangian 
have a generic $\cO(1)$ structure, such as the  up-type  
Yukawa coupling in Eq.~(\ref{eq:defZU}). In this framework 
the suppression of FCNCs arise by the rescaling the fermion fields 
necessary for the canonical normalization of the kinetic terms:
\beq
\label{eq:transf1}
Q_A \to Z_A Q_A~, \qquad   
\eeq
Starting from non-hierarchical bilinear structures,
\beq
\bar A  X_{AB} B ~, \qquad X_{AB}^{ij} =\cO(1)~,
\label{order1}
\eeq
the transformation into the canonical basis move the $Z_A$ into the 
effective couplings, with a corresponding  
suppression of the flavour-changing terms:
\beq
\label{eq:transf2}
 X^{ij}_{AB} \to \left(Z_A X_{AB} Z_B\right)^{ij}  ~\sim~ 
 z_A^{(i)}~z_B^{(j)}~.
\eeq
{ This illustrates a difference between hierarchical
wavefunctions,  and Froggatt-Neilson \cite{Froggatt:1978nt}
 type models \cite{Leurer:1992wg,Leurer:1993gy,Ramond:1993kv}.
 Both
may reproduce the observed Yukawa hierarchy, but a
simple Froggart-Neilson model \cite{Leurer:1992wg}, where
$ z_A^{(i)} \sim \epsilon^{Q_A^i}$ would give 
less suppression of FCNC processes:
$ X^{ij}_{AB} \to  \epsilon^{|Q_A^i - Q_B^j|}$.}

In the next section we analyse which conditions the $Z_A$ should
satisfy in order to provide a suppressions of FCNCs 
compatible with experimental data, while keeping the effective 
scale of New Physics in the TeV range.
 As we will show, in the quark sector these conditions are 
compatible with Eqs.~(\ref{eq:defZQD})--(\ref{eq:defZU}),
namely with a natural generation of the observed 
Yukawa couplings starting from generic $\cO(1)$ structures. 
The only exceptions are the kaon constraints from 
$\epsilon_K$ and  $\epsilon'/\epsilon$, which however 
can be fulfilled with a modest fine tuning ($\sim 10^{-1}$)
in the coupling of the effective operators.

\section{Bounds from quark FCNCs}
\label{sect:bounds}

\subsection{$\Delta F=2$ operators}
\unboldmath
As a first step we would like to constrain the parameters introduced 
in (\ref{parameterZ}) with the help of processes involving 
$\Delta F=2$ operators. 

There are in principle eight dimension-six four-quark operators 
that can contribute to down-type  $\Delta F=2$ processes \cite{Buras:2001ra}. 
However, restricting the attention to operators which preserve the 
$SU(2)_L \otimes U(1)_Y$ gauge symmetry, this number reduces to four:
\bea
O_{LL}^{ij}=&{\frac{1}{2}} {(\overline{ Q_L}^i \gamma_\mu Q^j_L)}^2~,  
\qquad\qquad O_{LR1}^{ij}=& (\overline{ Q_L}^i \gamma_\mu Q^j_L)(
\overline{D_R}^i \gamma_\mu D^j_R)~,\nonumber\\
 O_{RR}^{ij}=& \frac{1}{2}(\overline{ D_R}^i \gamma_\mu D^j_R)^2~, \qquad \qquad O_{LR2}^{ij}=& (\overline{ D_R}^i  Q^j_L)(\overline{Q_L}^i  D^j_R)~.
\eea
In order to derive model-independent bounds on the coupling 
of these operators, we introduce the following effective Hamiltonian 
(defined at the at the electroweak scale with 
canonically-normalized fields):
\bea
\mathcal{H}_{\rm eff} &=& 
 \left(C_{\rm SM}^{ij}+\frac{{X_{LL}^{ij}}}{\Lambda^2}\right)O_{LL}^{ij} +\frac{{X_{RR}^{ij}}}{\Lambda^2} O_{RR}^{ij}
+\frac{X_{LR1}^{ij}}{\Lambda^2}O_{LR1}^{ij} +\frac{X_{LR2}^{ij}}{\Lambda^2} O_{LR2}^{ij} + ~ {\rm h.c.} \\  \label{Heff}
&=& C_{LL}^{ij}(M_W) O_{LL}^{ij}
+C_{RR}^{ij}(M_W)O_{RR}^{ij}+C_{LR1}^{ij}(M_W)O_{LR1}^{ij}
+C_{LR2^{ij}}(M_W)O_{LR2}^{ij} + 
~{\rm h.c.}~, \quad \no
\eea
where 
\beq
\label{CSM}
C^{ij}_{\rm SM}= \frac{G_F^2}{2 \pi^2} M_W^2(V_{ti}^* V_{tj})^2 S_0(x_t)~.
\eeq
The amplitudes of the various $\Delta F=2$ processes 
can be calculated renormalizing the effective Hamiltonian (\ref{Heff}) 
at the relevant low scale $\mu$ 
(e.g.~$\mu\approx m_b$ for $\bar B^0-B^0$ mixing). 
The RGE running of the Wilson coefficients can be written as 
\beq\label{Cevolution}
\left(\begin{array}{c}
C_{LL}^{ij}(\mu)\\
 C_{RR}^{ij}  (\mu)\\
\vec C_{LR}^{ij} (\mu)
\end{array}
 \right)= \left(
\begin{array}{ccc}
 \eta_{LL}^{ij}(\mu, M_W) && \\
& \eta_{RR}^{ij}(\mu, M_W)&  \\
& &  \hat \eta_{LR}^{ij}(\mu, M_W)
\end{array}
 \right) \left(\begin{array}{c}
C_{LL}^{ij}(M_W)\\
 C_{RR}^{ij} (M_W) \\
\vec C_{LR}^{ij} (M_W)
\end{array}
 \right)~.
\eeq
Since QCD preserves chirality, there is 
no mixing between the $LL$, $RR$ and $LR$ sectors
and $\eta_{LL}^{ij}=\eta_{RR}^{ij}$.
The only non-trivial mixing occurs among the two $LR$ operators,
where $\hat \eta_{LR}^{ij}(\mu, M_W)$ is a $2\times 2$ matrix and
\beq
\vec C_{LR}^{ij}= \left(\begin{array}{c}
C_{LR1}^{ij}(\mu)  \\
C_{LR2}^{ij} (\mu)
\end{array}
 \right)=\hat \eta_{LR}^{ij}(\mu, M_W) 
\left(\begin{array}{c}
C_{LR1}^{ij}(M_W)  \\
C_{LR2}^{ij}(M_W)
\end{array}
 \right)~.
\eeq

The analytic formulae for these RGE factors 
as well as the relevant hadronic matrix elements
can be found in Ref.~\cite{Buras:2001ra}.
Following the approach given therein, we can 
express the $\Delta F=2$ amplitude for a generic 
neutral meson mixing as
\beq
\mathcal{A}^{ij} = \langle\overline M^0_{ij} 
|\mathcal{H}_{\rm eff}(\mu)| M^0_{ij} \rangle
\propto P_{LL}(C_{LL}+ C_{RR})+ P_{LR1}C_{LR1}+ P_{LR2} C_{LR2}~,
\label{firstamplitude}
\eeq
where the $P_A$ factors are 
appropriate combinations of RGE coefficients 
and hadronic matrix elements. Experimentally we have 
several precise constraints on this type of amplitudes:
both $|\mathcal{A}^{31}|$ and arg($\mathcal{A}^{31}$)
are constrained by $\Delta M_{B_d}$ and $S_{\psi K_S}$; 
$|\mathcal{A}^{32}|$ is constrained by $\Delta M_{B_s}$;
Im($\mathcal{A}^{12}$) is constrained by $\epsilon_K$.
In the following we will impose that the non-standard 
contribution is within $\pm$10$\%$ of  
the SM contribution, in magnitude, for all 
down-type $\Delta F=2$ mixing amplitudes.

Let us first analyze the $LL$ sector. Here the situation is quite simple 
since we can factorise the new-physics contribution as a correction 
to the SM Wilson coefficient:
\beq
\label{amplitude}
\left. \mathcal{A}^{ij} \right|_{LL}  
=\cA^{ij}_{\rm SM} \left(1 + \frac{ X_{LL}^{ij} 
}{(V_{ti}^* V_{tj})^2} \frac{F^2}{\Lambda^2}\right), 
\qquad
F=\left(\frac{2 \pi^2}{G_F^2 M_W^2 S_0(x_t)}\right)^{\frac{1}{2}}
\approx 3~{\rm TeV}~.
\eeq
{Since the effective scale of the SM contribution is 3 TeV, 
if new physics contributes via loops and is weakly interacting
(as the electroweak SM contribution),
taking $\Lambda \sim$  10 TeV corresponds to new masses  of the order of $3 M_Z$}.
Allowing for the amplitude (\ref{amplitude}) to vary from 
its SM values by at most $\pm$10$\%$, in magnitude, lead to 
\beq
\sqrt{|X_{LL}^{ij}|} <0.3\, |V_{ti}^* V_{tj}|\, \left(\frac{\Lambda}{F}\right)
 \quad < \ |V_{ti}||V_{tj}| \quad  {\rm for}~\Lambda < 10~{\rm TeV}~.
\eeq 
%%{ Recall that  $X_{LL}^{ij}$ is at the scale $m_W$.}
Expressing the flavour structure of the $X_{LL}^{ij}$ in terms 
of the corresponding hierarchical fermion wave functions, 
as in Eqs.~(\ref{eq:transf1})--(\ref{eq:transf2}), we find
\beq
\label{LLestimate}
\sqrt{|X_{LL}^{ij}|} \sim| z_Q^{(i)} z_Q^{(j)}| < |V_{ti}||V_{tj}|
\quad \to \quad   |z_Q^{(i)}| < |V_{ti}|~.
\eeq
Since $\eta_{LL}=\eta_{RR}$ and $\langle O_{LL}^{ij}\rangle=\langle O_{RR}^{ij}\rangle$, 
the constraint on the $RR$ operator is completely analog to the $LL$ one:
\beq\label{RRestimate}
\sqrt{|X_{RR}^{ij}|} \sim| z_D^{(i)} z_D^{(j)}| < |V_{ti}||V_{tj}|
\quad \to \quad   |z_D^{(i)}| < |V_{ti}|~.
\eeq

In the $LR$ sector the situation is slightly more complicated 
because of the different matrix elements involved. 
The $P_A$ factors introduced in Eq.~(\ref{firstamplitude}) can 
be decomposed as \cite{Buras:2001ra}:
\bea
P_{LL} &=&   \eta_{LL}(\mu, M_W)B_{LL}(\mu)~, \\
P_{LR1}&=&-  \hat\eta_{LR}(\mu, M_W)_{11}\left[B_{LR1}(\mu)\right]_{\rm eff}
 +\frac{3}{2}\hat\eta_{LR}(\mu, M_W)_{21}\left[B_{LR2}(\mu)\right]_{\rm eff}~, \\
P_{LR2}&=&-  \hat\eta_{LR}(\mu, M_W)_{12}\left[B_{LR1}(\mu)\right]_{\rm eff}
 +\frac{3}{2}\hat\eta_{LR}(\mu, M_W)_{22}\left[B_{LR2}(\mu)\right]_{\rm eff}~.
\eea
In the specific case of $\bar B^0-B^0$ we can further write 
\bea
&& \eta_{LL}(\mu, M_W)B_{LL}(\mu) = \eta_B \hat B_{B_q}~, \\
&& \left[B_{LRi}(\mu)\right]_{\rm eff}\ = \left(\frac{m_{B_q}}{m_b(\mu)+m_q(\mu)}\right)^2 B_{LRi}(\mu)~,
\eea
where $\hat B_{B_q}$ is the RGE invariant bag factor of the SM ($LL$) operator.
Using the
RGE factors in Ref.~\cite{Buras:2001ra, RGE} and the
$B_{LRi}(\mu)$ factors from lattice \cite{lattice,Bona:2007vi} leads to 
\bea
 P_{LL} = 0.7~,  \qquad P_{LR1}=-5.0~, \qquad P_{LR2}=6.3~,
\eea
with negligible differences between  $B_s$ and $B_d$ cases.
The contributions of the  $O_{LR1(2)}$  operator 
is thus enhanced by a factor $|P_{LR1(2)}/ P_{LL}| \approx 7 (9)$ 
compared to the SM one. Allowing  at most $\pm$10$\%$
corrections to the SM amplitude, the bounds derived 
from  $B_s$ and $B_d$ mixing are 
\bea
|X_{LR1}^{3j}| &\sim~| z_Q^{(3)}  z_Q^{(j)} z_D^{(3)}  z_D^{(j)} | & < ~0.2 ~|V_{tj}|^2~, \label{eq:BBmix1} \\
|X_{LR2}^{3j}| &\sim~| z_D^{(3)}  z_Q^{(j)} z_Q^{(3)}  z_D^{(j)} | & < ~0.1 ~ |V_{tj}|^2~, \label{eq:BBmix2}
\eea
where we have set $V_{tb}=1$ { and
$\Lambda < 10 $ TeV. These inequalities are satisfied,
with $\lambda_D^{ii}$ evaluated at $m_W$,} if 
we assume the hierarchy
\beq
z_Q^{(i)} z_D^{(i)} = (\lambda_D)_{ii}~,
\label{eq:ZdZq}
\eeq
which follows from the definition of $z_Q^{(i)}$ 
and $z_D^{(i)}$  in  Eq.~(\ref{eq:defZQD}).

Matrix elements and RGE factors lead to 
$P_{LR1(2)}$ substantially larger  
in the case of  $\bar K^0-K^0$
mixing~\cite{Buras:2001ra}:
\beq
P_{LL}=
0.5,\qquad P_{LR1}=-0.7 \times 10^2~,\qquad P_{LR2}=1.1 \times 10^2~.
\eeq
Proceeding as in the  $\bar B^0-B^0$ case, this implies 
the stringent bound
\beq
|X_{LR2}^{21}|  \sim | z_Q^{(2)}  z_Q^{(1)} z_D^{(2)}  z_D^{(1)} |  < ~0.004~|V_{ts}^* V_{td}|^2 
\approx   0.6 \times 10^{-9}~, \label{KKLR2}
\eeq
and similarly for $|X_{LR1}^{21}|$. 
In such case, using Eq.~(\ref{eq:ZdZq}) the bound is not fulfilled 
by about one order of magnitude:\footnote{~Note that in the case of
$\bar K^0-K^0$ mixing we don't have a stringent experimental 
constraint on the modulo of the amplitude (given the large
long-distance contributions to $\Delta M_K$) but only on its 
imaginary part (thanks to $\epsilon_K$): thus the 
order of magnitude disagreement concerns only the 
CPV part of the $\Delta S=2$ amplitude.}
\beq
| z_Q^{(2)}  z_Q^{(1)} z_D^{(2)}  z_D^{(1)} | \sim 
\frac{ m_d m_s }{v^2}   \approx 1 \times  10^{-8}~. 
\eeq
{ This result is qualitatively similar to the 
conclusion of Ref.~\cite{Bona:2007vi}, where $\epsilon_K$ 
has been identified as the most significant $\Delta F=2$ 
constraint on (non-MFV) models where
NP couples dominantly to the third generation.}

\subsection{$\Delta F=1$ operators}

Taking into account the analysis of the  $\Delta F=2$ sector, 
it is clear that  $\Delta F=1$ operators with a $LL$ (or $RR$) structure,
such as $\overline{ Q_L}^i \gamma^\mu {Q_L}^j H^\dagger D_\mu H$ 
do not represent a serious problem. The corresponding 
constraints are equivalent to those derived from $LL$ and $RR$
$\Delta F=2$ operators, which are naturally fulfilled.
On the other hand, 
a potentially interesting class of new constraints in the $\Delta F=1$
sector arises by magnetic and chromomagnetic operators:
\beq
O^{ij}_{RL\gamma} =e H^\dagger \overline{D_R}^i \sigma^{\mu \nu} Q^j_L F_{\mu \nu}~, \qquad 
O^{ij}_{RLg} =g_s H^\dagger \overline{ D_R}^i \sigma^{\mu \nu} T^a Q^j_L G^a_{\mu \nu}~. 
\eeq

In the case of the $O^{Fij}_{RL\gamma}$ operators the most 
significant constraint is derived from  $B\to X_s \gamma$.
The leading effective Hamiltonian at the electroweak scale 
can be written as
\beq
\mathcal{H}_{\rm eff} = \left(C^{32}_{\rm SM} + \frac{X^{32}_{RL\gamma}}{\Lambda^2} 
\right) O^{32}_{RL\gamma} +  \frac{X^{23}_{RL\gamma}}{\Lambda^2} O^{23}_{RL\gamma}~ + ~
{\rm h.c.}~, 
\eeq 
where
\beq
 C^{32}_{\rm SM} = -\frac{G_F}{ 4\pi^2 \sqrt{2} } \lambda_b V_{ts}^* V_{td} C_7^{\rm SM}(M_W)~,
\eeq
and $ C_7^{\rm SM}(M_W^2) \approx -0.3$ is defined as in Ref.~\cite{Misiak:2006zs}.
The contribution of  $O^{32}_{RL\gamma}$ operator, which encodes also
the SM contribution, can simply be taken into account by a shift of 
$C_7^{\rm SM}(M_W^2)$ at the electroweak scale:
\beq
\frac{ \delta C_7(M_W)}{C^{\rm SM}_7(M_W)} 
= \frac{ X^{32}_{RL\gamma}}{ V_{ts}^* V_{tb}} \frac{1}{\lambda_b}\frac{F^2_B  }{\Lambda^2}~,
\qquad 
F_B=\left(-\frac{4 \pi^2 \sqrt{2}}{C_7^{\rm SM} G_F}\right)^{\frac{1}{2}}\approx 5 \,{\rm TeV}~.
\eeq
Using the approximate expression~\cite{Misiak:2006zs,Misiak}
\beq
\frac{\cB(B \to X_s \gamma)}{\cB(B \to X_s \gamma)_{\rm SM}}\approx 1+2.9\times \delta C_7(M_W)~, 
\eeq
and allowing for a 15\% departure from the SM value, leads to the bound
({ for $\Lambda < 10 $ TeV}):
\beq
|X^{32}_{RL\gamma}| \sim 
|z_D^{(3)} z_Q^{(2)}| <\,0.5 \, |V_{ts}^* V_{tb}|\lambda_b \frac{\Lambda^2}{F_B^2}
<  1.2 \times 10^{-3}~.
\label{eq:bsgb}
\eeq
Employing the hierarchy~(\ref{eq:ZdZq}) and assuming $|z_Q^{(i)}| \sim |V_{ti}|$
(i.e.~saturating the constraint (\ref{LLestimate}) from $\Delta F=2$ $LL$ operators), 
this bound is naturally fulfilled:
\beq
|z_D^{(3)}z_Q^{(2)}|\sim |\lambda_b V_{ts}/V_{tb} | \approx  7\times 10^{-4}~.
\eeq 
Employing the same hierarchy, the coupling of the 
$O^{23}_{RL\gamma}$ operator is substantially larger:
$|X^{23}_{RL\gamma}| \sim |z_D^{(2)} z_Q^{(3)}| \sim  |\lambda_s V_{tb}/V_{ts} |
\approx 8\times 10^{-3}$. However, since this operator does not interfere 
with the SM contribution, the bound on $X^{23}_{RL\gamma}$ 
is weaker with respect to one in Eq.~(\ref{eq:bsgb}): 
$|X^{23}_{RL\gamma}| < 6\times 10^{-3}$. 
We then conclude that the constraints from $B\to X_s \gamma$ are 
essentially fulfilled without fine tuning. 

Similarly to the $\Delta F=2$ sector, the most serious problems arise from $K$ 
physics. Here the most significant constraints are obtained from the 
possible impact in $\epsilon^\prime/\epsilon$
of the chromomagnetic operators. The contribution 
of $\Delta I=1/2$ operators (such as $O^{12}_{RLg}$ and $O^{21}_{RLg}$) 
to $\epsilon^\prime/\epsilon$  can be generally written as 
\beq
\delta {\rm Re}\left(\frac{\epsilon^\prime}{\epsilon}\right) \approx  
\frac{ \omega}{ \sqrt{2}  |\epsilon|  {\rm Re} \cA_0  } \times \delta {\rm Im} \cA_0~,
\eeq
where $\cA_I = \cA(K^0 \to 2\pi_{I})$ and $\omega = |A_2/A_0| \approx 1/22$~. 
In the specific case of the chromomagnetic operators, following 
Ref.~\cite{Colangelo:1999kr, Buras}, we have 
\bea
|\delta {\rm Im} \cA_0 | &=& 
  \frac{| {\rm Im}(X^{12}_{RLg} - X^{21}_{RLg})| v }{  \Lambda^2 } ~\eta_G~ 
 \langle 2\pi_{I=0} | g_s \bar s_R (\sigma\cdot G) d_L | K^0 \rangle \no \\
 &=& \frac{| {\rm Im}(X^{12}_{RLg} - X^{21}_{RLg})| }{ \Lambda^2 } ~ \eta_G B_G 
\sqrt{\frac{3}{2}} \frac{11}{2} \frac{ m_\pi^2 m_K^2 }{ F_\pi \lambda_s}~,
\eea
where $X^{12(21)}_{RLg}$ are the couplings of the effective operators at
the electroweak scale (defined in analogy to the $X^{12(21)}_{RL\gamma}$),
 $\eta_G$ is the  RGE factor, and  $B_G$ denote 
the bag parameter of the hadronic matrix element. Using the numerical values 
in Ref.~\cite{Colangelo:1999kr} and imposing that the extra contribution to 
$\epsilon^\prime/\epsilon$ do not exceed $10^{-3}$ leads to the following 
bound:
\beq
\frac{ |{\rm Im}(X^{12}_{RLg} - X^{21}_{RLg})| }{ \lambda_s}~ < ~
10^{-2} \left( \frac{\Lambda}{ 10~{\rm TeV} } \right)^2~.
\eeq
Using, as in the previous cases, 
the hierarchy~(\ref{eq:ZdZq}) and $|z_Q^{(i)}| \sim |V_{ti}|$, 
we obtain 
\beq
\left|\frac{ X^{21}_{RLg} }{\lambda_s} \right| \sim \left|\frac { z_D^{(2)}z_Q^{(1)} }{\lambda_s} \right|
 \sim \left| \frac{V_{td}}{V_{ts}} \right| \approx 0.2~, \qquad 
\left|\frac{ X^{12}_{RLg} }{\lambda_s} \right| \sim \left|\frac { z_D^{(1)}z_Q^{(2)} }{\lambda_s} \right|
 \sim \left| \frac{ \lambda_d V_{ts}}{\lambda_s V_{td}} \right| \approx 0.3~.
\eeq
Similarly to the case of $\bar K^0-K^0$ mixing,
also in this case the suppression implied by the $z_A^{(i)}$ leads 
to a natural size about one order of magnitude larger 
than what is needed to fulfill the experimental constraints.
In close analogy with the $\Delta S=2$ case, the problem 
arises only from the imaginary (CP-violating) part of 
the amplitude.

\section{Operators involving lepton fields}
\label{sect:leptons}

The approach we have followed so far in the quark sector 
can easily be exported also to the lepton sector. 
Introducing the diagonal matrices 
$Z_L$ and $Z_E$, such that 
\beq
Z_L^{-1}\lambda_E Z_E^{-1} \simeq I~, \qquad 
\lambda_E = \frac{1}{v} {\rm diag}(m_e, m_\mu, m_\tau)~,
\label{guessleptons}
\eeq 
we proceed investigating the impact of the transformation 
\beq
L_L\to Z_L L_L~ \qquad  E_R\to Z_E E_R~,
\label{eq:leptons}
\eeq
onto operators involving lepton fields.
A major difference with respect to the quark sector is that
lepton flavour is conserved  in the SM ($d=4$) Lagrangian.
The only observed lepton-flavour changing transitions
are in neutrino oscillations, whereas lepton-flavour violating (LFV) transitions 
of charged leptons are severely constrained by experiments. 

Before analysing the efficiency  of the transformation  (\ref{eq:leptons})
in suppressing LFV processes, it is worth stressing that it has a non-trivial impact also 
in quark FCNC transitions. Indeed four-fermion
operators with a leptonic current, 
such as
\beq
\overline{Q_L}^i \gamma^\mu Q_L^j \overline{ L_L}^k \gamma_\mu L_L^\ell~, \qquad  
\overline{ D_R}^i Q_L^j \overline{ L_L}^k  E_R^\ell ~ ,
\label{eq:QLFV}
\eeq
receive lepton suppression factors 
in addition to those from the quark wave functions
(the coefficients are respectively 
$\sim z_{Q}^{(i)} z_Q^{(j)}  z_L^{(k)} z_L^{( \ell)}$ and 
$\sim z_Q^{(j)} z_D^{(i)}  z_L^{(k)} z_E^{( \ell)}$).
 This implies that 
such operators are totally negligible.
Their contributions to
lepton-flavour conserving processes, 
such as $B(K)\to \ell^+\ell^-$ or
$B(K) \rightarrow \pi \ell^+\ell^-$,
are well below the size  
expected in the MFV framework.
Similarly, their  contributions to
neutral-current processes which violate both 
quark and lepton flavour, such as 
$B_0 \rightarrow \tau \bar{\mu} $ or  $K_L^0 \rightarrow \mu \bar{e}$,
are well below the current experimental sensitivity.

\subsection{Bounds from LFV processes}

In the lepton sector the most stringent constraints are 
on $\Delta F = 1$ transitions among the first two generations
($\mu \rightarrow e \gamma$, $\mu \rightarrow e \bar{e} e$
and $\mu \to e$ conversion on nuclei).  $\Delta F = 2$ processes,
such as muonium-anti-muonium conversion \cite{Willmann:1998gd},
are less restrictive.  In our scenario the largest rates for $\Delta F = 1$ 
processes arise from higher-dimensional operators 
bilinear in the lepton fields (suppressed only by two 
$z_{L,E}$ factors) which induce
$\mu e \gamma$ and $ \mu e Z$ effective interactions. 
So  we focus on  electroweak
dipole operators such as
\beq
\label{WHAT}
O_{RL1}^{ij} =g^\prime H^\dagger \overline{E_R}^i \sigma^{\mu \nu}L^j_L B_{\mu \nu}~,\qquad 
O_{RL2}^{ij} =g H^\dagger \overline{ E_R}^i \sigma^{\mu \nu} \tau^a L^j_L W^a_{\mu \nu}~.
\eeq
and operators contributing
to the $\ell \ell^\prime Z$ vertex\footnote{~The complete basis of LFV 
operators can be found for instance in Ref.~\cite{LFVops}}
\beq
% \delta \mathcal{L}=\frac{1}{\Lambda^2} \left(
O_{LL}^{ij} = \overline{L_L}^i \gamma^\mu L^j_L H^\dagger D_\mu H~,\qquad 
O_{RR}^{ij} = \overline{E_R}^i \gamma^\mu E^j_R H^\dagger D_\mu H~. 
\label{eq:Zvertex}
\eeq

We start analysing the constraints from the radiative LFV decays, 
which are sensitive to dipole operators only and 
which turn out to be the most significant processes.  
Introducing the 
effective Lagrangian 
\beq
\mathcal{L}= \frac{1}{\Lambda^2}\sum  X_{RLx}^{ij} O_{RLx}^{ij} ~+~{\rm h.c.}
\eeq
the radiative  branching ratios can  be written 
as\footnote{~We have neglected the helicity-suppressed interference 
term between $X_{RL\gamma}^{ij}$ and $X_{RL\gamma}^{ji}$} \cite{Hisano:1995cp}
\bea
\cB(l_i \to l_j \gamma)
= \frac{\Gamma(l_i \to l_j \gamma)}{\Gamma(l_i \to l_j \nu \bar{\nu})}
 \cB(l_i \to l_j \nu \bar{\nu}) 
= \frac{ 192 \,\pi^3\, \alpha }{G_F^2\Lambda^4 }\frac{1}{ (\lambda_E)_{ii}^2} \left[ |X_{RL\gamma}^{ij}|^2 + |X_{RL\gamma}^{ji}|^2 
\right] b_{ij} ~,
\eea
where $X_{RL\gamma}^{ij}=X_{RL2}^{ij}-X_{RL1}^{ij}$ 
and $b_{ij} = \cB(l_i \to l_j \nu \bar{\nu}) $,  
$\{b_{\mu e},b_{\tau e},b_{\tau \mu}\}=\{1.0,0.178,0.173\}$.
We can already see that these branching ratios
tend to be large in our scenario:
the $z_L^{(i)} z_E^{(j)}$ suppression is 
partially compensated by the $1/(\lambda_E)^2_{ii}$ term 
(which appears because of the normalization 
to $\Gamma(l_i \to l_j \nu \bar{\nu})\propto m_i^5$)
and if the new physics gives 
$l_i \to l_j \gamma$  via loops, the associated
 $1/16 \pi$ (which is absorbed in $\Lambda^2$) 
is compensated by the three body final
state phase space of $l_i \to l_j \nu \bar{\nu}$.

The decomposition 
$ X_{RL}^{ij} \sim z_E^{(i)} z_L^{(j)}$
% and fixing the $z_E^{(i)}$ by the condition 
% $z_E^{(i)}z_L^{(i)}= (\lambda_E)_{ii}$
leads to
\bea
\cB(\mu \to e \gamma) &\approx& 1.2 \times 10^{-11}
\left(\frac{130~\tev}{\Lambda}\right)^4   \left( 
\frac{|  z_L^{(1)} z_E^{(2)}|^2 }
{ 2(\lambda_E)_{22}(\lambda_E)_{11}} 
  +  \frac{|  z_E^{(1)} z_L^{(2)}|^2 }
{ 2(\lambda_E)_{22}(\lambda_E)_{11}} 
\right)~,  \label{megbd}
  \\
\cB(\tau \to e \gamma)&\approx& 1.1 \times 10^{-7}
\left(\frac{4.3~\tev}{\Lambda}\right)^4 
% \frac{1}{ 2(\lambda_E)_{11}(\lambda_E)_{33}} 
%\left( \left|  z_L^{(1)} z_E^{(3)}\right|^2 
%  +  \left|  z_E^{(3)} z_L^{(1)}\right|^2 
%\right)~,
\left( 
\frac{|  z_L^{(1)} z_E^{(3)}|^2 }
{ 2(\lambda_E)_{11}(\lambda_E)_{33}} 
  +  \frac{|  z_E^{(3)} z_L^{(1)}|^2 }
{ 2(\lambda_E)_{33}(\lambda_E)_{11}} 
\right)~,
% \left( \left| \frac{z_L^{(1)}}{z_L^{(3)}} \right|^2 
%  +  \left| \frac{z_{E}^{(1)}}{z_E^{(3)}} \right|^2 \right)~,  
  \\
\cB(\tau \to \mu \gamma)&\approx& 4.5 \times 10^{-8}
 \left(\frac{20~\tev}{\Lambda} \right)^4
\left( 
\frac{|  z_L^{(2)} z_E^{(3)}|^2 }
{ 2(\lambda_E)_{22}(\lambda_E)_{33}} 
  +  \frac{|  z_E^{(3)} z_L^{(2)}|^2 }
{ 2(\lambda_E)_{33}(\lambda_E)_{22}} 
\right)~,
% \left( \left| \frac{z_L^{(2)}}{z_L^{(3)}} \right|^2 
%  +  \left| \frac{z_{E}^{(2)}}{z_E^{(3)}} \right|^2 \right)~,  
\eea
where the scale for  each mode has been  chosen such that the
  branching ratio is close  to its current experimental 
bound~\cite{PDG,Hayasaka:2007vc,Aubert:2005wa}.

%Contrary to the quark sector, in this case we do not have 
%stringent independent bounds on the $z^{(i)}_L$,
%therefore we cannot immediately translate these
%expressions into bounds on $\Lambda$. 
Using  a CKM-type ansatz ($z_L^{(3)}\sim 1$, $z_L^{(2)}\sim \lambda^2$, 
$z_L^{(1)}\sim \lambda^3$, with  $\lambda \sim 0.2$),  for $\Lambda < 10~\tev$
both $\tau \to \mu \gamma$
and  $\tau \to e \gamma$
are within their experimental bounds.
On the contrary,
$\mu\to e \gamma$ exceeds its bound by six (!) orders 
of magnitude. The problem of  $\mu\to e \gamma$ 
persists with any reasonable ansatz 
(such as the ``democratic'' assignment
$z_L^{(i)}/z_L^{(j)} \sim z_E^{(i)}/z_E^{(j)} \sim (m_i/m_j)^{1/2}$).
We thus conclude that 
either the scale of the LFV operators is pushed well 
above $10~\tev$ or, equivalently, 
the corresponding couplings are suppressed by 
several orders of magnitude compared to their 
naive expectation in this framework.\footnote{~For instance,
if the dipole operator is generated only via an effective 
four-lepton interaction (with two lepton lines 
closed into a loop), its coupling  receives an extra 
suppression factor which allow to set $\Lambda \sim 10 $ TeV.
Similarly,  dipole operators are dynamically suppressed in the 
the RS scenario considered 
in Ref.~\cite{Kitano:2000wr}, where the 
new degrees of freedom are only  vector-like.}

Similar (slightly less stringent) bounds on the dipole  operators 
are obtained from their contributions to $\mu \to e$ 
conversion in nuclei and $\mu \rightarrow 3 e$.
We finally briefly consider the $LL$ and $RR$ operators
in Eq.~(\ref{eq:Zvertex}). After electroweak symmetry breaking, 
and integrating out the heavy $Z$ field, these
give rise to four-fermion operators of the type $\bar{f}^k \gamma^\mu
f^k \overline{L_L}^i \gamma_\mu L_L^j$, where the $f^k$ can be any of the SM
fermions.  These contribute to $\ell_i \to 3 \ell$ decays , $\mu \to e$
conversion,  and other previously discussed
quark and lepton operators. For $\Lambda \sim 10$ TeV, 
and with 
``democratic'' assignment
$X_{LL,RR}^{ij}\sim \sqrt{(\lambda_{E})_{ii} (\lambda_{E})_{jj}}$,
the expected rate are all below the experimental bounds.
%We have explicitely checked that  their contribution to  $\mu \to e$ 
%conversion is within the present experimental bounds
%using the ``democratic'' assignment
%$X_{LL,RR}^{ij}\sim \sqrt{(\lambda_{E})_{ii} (\lambda_{E})_{jj}}$.

\section{Discussion and comparison to MFV}
\label{sect:comparison}

The analysis of the previous sections can be summarized as follows.
In the quark sector the hierarchy of the fermion 
kinetic terms necessary
 to naturally reproduce both 
Yukawa structures and suppress dimension-six $LL$ operators is 
% for a natural size of 
%both Yukawa structures and dimension-six $LL$ operators is 
\beq
z_Q^{(i)}\sim V_{ti}
 \sim \mathcal{O}(1), \mathcal{O}(\lambda^2), \mathcal{O}(\lambda^3),
\qquad z_D^{(i)} \sim \frac{(\lambda_D)_{ii}}{V_{ti}}
\sim  \mathcal{O}(\lambda^2)~, \mathcal{O}(\lambda^3), \mathcal{O}(\lambda^4)~.
\label{eq:hierf}
\eeq
Employing this hierarchy, most of the predictions from 
other $\Delta F=2$ and $\Delta F=1$ dimension-six FCNC 
operators  fall  within the experimental bounds without fine 
tuning, i.e.~assuming generic $\cO(1)$ couplings in the 
basis where the kinetic terms are  hierarchical, and  
an effective scale of new physics  $ \sim 10~\tev$.
The only exceptions are the 
$LR$ operators contributing to $\epsilon_K$ 
and  $\epsilon'/\epsilon$, which 
can be sufficiently suppressed with a modest fine tuning
of the hierarchies and $\cO(1)$ coefficients.\footnote{~To 
estimate the amount of fine tuning, we assume the $\cO (1)$ 
coefficients $c$ to be distributed as
$P(|c|) \propto {\rm exp} \{ -( {\rm ln}|c|)^2/2 \}$.
Under this assumption the coefficients of the FCNC
operators are obtained via a maximum likelihood
analysis and we find a $\sim 15 \%$ probability
to be consistent with the experimental bounds.}
This scenario predicts that new physics in  $\epsilon_K$ 
and  $\epsilon'/\epsilon$ is just around the corner:
a conspiracy %of  $\cO (1)$ factors  and hierarchies
to suppress the $LR$ operator coefficients by
another order of magnitude 
would require a too severe fine tuning
(it occurs with a probability $\lsim  0.5 \%$).

Given the consistency of this general framework in 
suppressing quark FCNC amplitudes, it is useful to compare 
it  with the more precise predictions of the MFV framework
(from which the fuzzy $ \cO(1)$ factors are absent).  
In Table \ref{tab:MFV} we list quark bilinears and their
suppression factors, as expected   
within MFV 
and within the framework with hierarchical fermion profiles. 
The differences arise for $RR$ and $LR$ operators. 
Within the MFV hypothesis these are strongly suppressed by one or two 
powers of the down-type Yukawa coupling. However, such as suppression 
is not necessary for the description  of nature (especially in the
case of $RR$ operators) and is partially removed in the 
scenario with hierarchical fermion profiles.
{ Note that, while in the MFV framework is possible 
to enhance the overall normalization of the down-type Yukawa 
couplings, considering two Higgs fields and a large vev
ratio ($\tan\beta= \langle H_U \rangle /\langle H_D \rangle \gg 1$),
this is not possible in the scenario with hierarchical 
kinetic terms. In the latter case a possible $\tan\beta$
enhancement of the Yukawa couplings would produce a 
strong tension with data in the $LR$ sector.}

\begin{table}[t]
\begin{center}
\begin{tabular}{|c|c|c|c|c|} \hline 
Quark Bilinears & \multicolumn{2}{|c|}{MFV} &  \multicolumn{2}{|c|}{Hierarchical Kinetic Terms} \\
                & {\footnotesize parametric size} & {\footnotesize comparison with exp.}
                & {\footnotesize parametric size} & {\footnotesize comparison with exp.} \\
\hline
$L^i L^j$  &  $V^*_{ti} V_{tj}$ & {\footnotesize close to experiment} & $V^*_{ti} V_{tj}$   & 
{\footnotesize same size as MFV}\\ 
\hline
$L^i R^j$ &  $(\lambda_D)_{ii} V^*_{ti} V_{tj}$ & {\footnotesize negligible}& 
 $(\lambda_D)_{ii} \dis\frac{V_{tj}}{ V_{ti}}$& {\footnotesize can exceed exp. bounds} \\
\hline  
$R^i R^j$ &  $(\lambda_D)_{ii} V^*_{ti} V_{tj}(\lambda_D)_{jj}$  & {\footnotesize negligible}&  $\dis\frac{(\lambda_D)_{ii}(\lambda_D)_{jj}}{V^*_{ti} V_{tj}}$
& {\footnotesize comparable to} $L^i L^j$\\
\hline
\end{tabular}
\end{center}
\caption{\label{tab:MFV} Comparison of the parametrical suppression factors 
of the various quark bilinears, both in the MFV framework and 
in the general scenario with  hierarchical kinetic terms.}
\end{table}

The situation of the lepton sector is  more
problematic. The constraints on helicity conserving ($LL$ and $RR$)
LFV operators are satisfied assuming an hierarchy of the type 
(\ref{eq:hierf}) for $z^{(i)}_L$ and  $z^{(i)}_E$ 
(with $\lambda_D \to \lambda_E$) or, equivalently, the 
democratic assignment $z_{L,E}^{(i)} \sim \sqrt{(\lambda_{E})_{ii}}$.
However, the constraints on $LR$ operators contributing to 
$\mu \to e\gamma$  and $\mu\to e$ conversion, 
require an effective scale in the $100~\tev$ range. 
It is therefore difficult to make predictions: if the LFV rates are
suppressed because the new physics scale $ \Lambda$  is
high in the lepton sector, then $\mu\to e\gamma$  could 
be close to its present exclusion bound, $\cB(\mu \to3 e)$ 
and the rate for $\mu \to e$ conversion are suppressed 
by $\cO(\alpha)$ with respect to  $\cB(\mu\to e\gamma)$,
and $\tau$ LFV decays are beyond the reach of future 
facilities. However, these predictions do not hold 
if LFV dipole operators 
are suppressed  by some specific dynamical mechanism. In the latter case we cannot exclude 
scenarios where the $\tau\to \mu\gamma$ rate is close to its present exclusion bound. 
 
Given the important role of dipole operators in this framework, 
it is worth to look at the bounds derived from the corresponding 
flavour-diagonal partners contributing to anomalous-magnetic
and electric-dipole moments, both in the quark and in the lepton sector.
As far as anomalous-magnetic moments are concerned, the most significant 
constraint comes from $(g-2)_\mu$. Here we could solve the current 
discrepancy~\cite{Bennett:2006fi} only for  
$\Lambda \sim  2-3 $ TeV (setting $ z_E^{(2)} z_L^{(2)} \sim (\lambda_E)_{22}$),
a scale which is far too too low 
compared to the $\mu \to e \gamma$ bound. We thus conclude that 
there is no significant contribution to  $(g-2)_\mu$ in this framework. 
On the other hand, stringent  bounds on $\Lambda$ (in the  $\sim 100$~TeV
range) are imposed by the electron and the neutron electric-dipole moments
(assuming $ z_{E(D)}^{(1)} z_{L(Q)}^{(1)} \sim (\lambda_{E(D)})_{11}$
and $\cO(1)$ flavour-diagonal CP-violating phases). 
Being flavour-conserving and CP-violating, 
the coupling of these operators could easily be suppressed 
by independent mechanisms, such as an approximate
CP invariance in the flavour-conserving part of the Lagrangian.
However, the fact that these bounds are close to those derived from 
$\mu\to e\gamma$ can also be interpreted as a further indication 
of a common dynamical suppression mechanism of dipole-type operators.

\section{Conclusions}

The absence of deviations from the SM  
in the flavour sector points toward extensions 
of the model with highly non-generic flavour structures.
In this paper we have investigated
the viability of models  where the suppression of 
flavour-changing transitions is not attributed to 
a specific flavour symmetry, but it arises only from 
appropriate hierarchies in the kinetic terms.
A generic framework which could occur in models 
with extra dimensions.
   
We have considered in particular the class of scenarios 
where the Yukawa matrices and the dimensionless 
flavour-changing couplings of the higher-dimensional operators 
do not exhibit a specific flavour structure (i.e.~have 
generic $\cO(1)$ entries) in a basis where the kinetic 
terms are hierarchical. Despite its simplicity,
this construction is sufficient to suppress to a level consistent 
with experiments all the flavour-changing 
transitions in the quark sector, assuming a scale 
of new physics in the TeV range. The only two observables 
where a mild tuning of the  $\cO(1)$ coefficient is needed
are the CP-violating parameters of the neutral kaon system:
within this scenarios new physics effects 
in  $\epsilon_K$ and  $\epsilon'/\epsilon$ should be detectable
with improved control on the corresponding  
SM predictions.

The most serious challenge to this class of models appears 
in the lepton sector, thanks to the stringent bounds on 
$\mu\to e$ transitions. The latter require either an heavy  
effective scale of LFV ($\Lambda_{\rm LFV} \gsim 100$ TeV)
or an independent dynamical suppression mechanism for 
dipole-type operators.

\subsection*{Acknowledgments}
We would like to thank Andrzej J.~Buras for informative discussions,
{ Kastubh Agashe for careful reading,  
and  Yossi Nir for thoughtful comments}. 
The work of S.U.~has been supported the Bundesministerium f\"ur Bildung und Forschung under the contract 05HT6WOA. 
The work of G.I.~is supported in part by the EU Contract No.~MRTN-CT-2006-035482 "FLAVIAnet".
S.D.~and S.U.~acknowledge the  hospitality  of the LNF  Spring Institute 2007.


\begin{thebibliography}{999}

{\footnotesize 

\bibitem{MFV0}
%\cite{Chivukula:1987py}
%\bibitem{Chivukula:1987py}
  R.~S.~Chivukula and H.~Georgi,
  %``Composite Technicolor Standard Model,''
  Phys.\ Lett.\  B {\bf 188} (1987) 99;
  %%CITATION = PHLTA,B188,99;%%
%\cite{Hall:1990ac}
%\bibitem{Hall:1990ac}
  L.~J.~Hall and L.~Randall,
  %``Weak scale effective supersymmetry,''
  Phys.\ Rev.\ Lett.\  {\bf 65} (1990) 2939.
  %%CITATION = PRLTA,65,2939;%%


%\cite{D'Ambrosio:2002ex}
\bibitem{D'Ambrosio:2002ex}
  G.~D'Ambrosio, G.~F.~Giudice, G.~Isidori and A.~Strumia,
  %``Minimal flavour violation: An effective field theory approach,''
  Nucl.\ Phys.\  B {\bf 645} (2002) 155
  [arXiv:hep-ph/0207036].
  %%CITATION = NUPHA,B645,155;%%



%\cite{Cirigliano:2005ck}
\bibitem{Cirigliano:2005ck}
  V.~Cirigliano, B.~Grinstein, G.~Isidori and M.~B.~Wise,
  %``Minimal flavor violation in the lepton sector,''
  Nucl.\ Phys.\  B {\bf 728} (2005) 121
  [arXiv:hep-ph/0507001].
  %%CITATION = NUPHA,B728,121;%%



%\cite{Buras:2000dm}
\bibitem{Buras:2000dm}
  A.~J.~Buras, P.~Gambino, M.~Gorbahn, S.~Jager and L.~Silvestrini,
  %``Universal unitarity triangle and physics beyond the standard model,''
  Phys.\ Lett.\  B {\bf 500}, 161 (2001)
  [arXiv:hep-ph/0007085].
  %%CITATION = PHLTA,B500,161;%%


\bibitem{Bona:2007vi}
  M.~Bona {\it et al.}  [UTfit Collaboration],
  %``Model-independent constraints on Delta F=2 operators and the scale of New
  %Physics,''
  arXiv:0707.0636 [hep-ph].
  %%CITATION = ARXIV:0707.0636;%%


\bibitem{Agashe:2005hk}
  K.~Agashe, M.~Papucci, G.~Perez and D.~Pirjol,
  %``Next to minimal flavor violation,''
  arXiv:hep-ph/0509117.
  %%CITATION = HEP-PH/0509117;%%


\bibitem{Davidson:2006bd}
  S.~Davidson and F.~Palorini,
  %``Various definitions of minimal flavour violation for leptons,''
  Phys.\ Lett.\  B {\bf 642} (2006) 72
  [arXiv:hep-ph/0607329].
  %%CITATION = PHLTA,B642,72;%%

\bibitem{Grinstein:2006cg}
  B.~Grinstein, V.~Cirigliano, G.~Isidori and M.~B.~Wise,
  %``Grand unification and the principle of minimal flavor violation,''
  Nucl.\ Phys.\  B {\bf 763} (2007) 35
  [arXiv:hep-ph/0608123].
  %%CITATION = NUPHA,B763,35;%%


\bibitem{Feldmann:2006jk}
  T.~Feldmann and T.~Mannel,
  %``Minimal flavour violation and beyond,''
  JHEP {\bf 0702} (2007) 067
  [arXiv:hep-ph/0611095].
  %%CITATION = JHEPA,0702,067;%%



\bibitem{BarShalom:2007pw}
  S.~Bar-Shalom and A.~Rajaraman,
  %``Models and Phenomenology of Maximal Flavor Violation,''
  arXiv:0711.3193 [hep-ph].
  %%CITATION = ARXIV:0711.3193;%%


%\cite{ArkaniHamed:1999dc}
\bibitem{ArkaniHamed:1999dc}
  N.~Arkani-Hamed and M.~Schmaltz,
  %``Hierarchies without symmetries from extra dimensions,''
  Phys.\ Rev.\  D {\bf 61}, 033005 (2000)
  [arXiv:hep-ph/9903417];
  %%CITATION = PHRVA,D61,033005;%%
  Y.~Grossman and M.~Neubert,
  %``Neutrino masses and mixings in non-factorizable geometry,''
  Phys.\ Lett.\  B {\bf 474} (2000) 361
  [arXiv:hep-ph/9912408]. 
  %%CITATION = PHLTA,B474,361;%%

\bibitem{Gherghetta}
  T.~Gherghetta and A.~Pomarol,
  %``Bulk fields and supersymmetry in a slice of AdS,''
  Nucl.\ Phys.\  B {\bf 586} (2000) 141
  [arXiv:hep-ph/0003129].
  %%CITATION = NUPHA,B586,141;%%


%\cite{Randall:1999ee}
\bibitem{Randall:1999ee}
  L.~Randall and R.~Sundrum,
  %``A large mass hierarchy from a small extra dimension,''
  Phys.\ Rev.\ Lett.\  {\bf 83}, 3370 (1999)
  [arXiv:hep-ph/9905221].
  %%CITATION = PRLTA,83,3370;%%
  

\bibitem{Kitano:2000wr}
  R.~Kitano,
  %``Lepton flavor violation in the Randall-Sundrum model with bulk
  %neutrinos,''
  Phys.\ Lett.\  B {\bf 481} (2000) 39
  [arXiv:hep-ph/0002279].
  %%CITATION = PHLTA,B481,39;%%



\bibitem{Huberetal}
  S.~J.~Huber,
  %``Flavor violation and warped geometry,''
  Nucl.\ Phys.\  B {\bf 666} (2003) 269
  [arXiv:hep-ph/0303183];
  %%CITATION = NUPHA,B666,269;%%
  G.~Burdman,
  %``Flavor violation in warped extra dimensions and CP asymmetries in B
  %decays,''
  Phys.\ Lett.\  B {\bf 590} (2004) 86
  [arXiv:hep-ph/0310144];
  %%CITATION = PHLTA,B590,86;%%
  S.~Khalil and R.~Mohapatra,
  %``Flavor violation and extra dimensions,''
  Nucl.\ Phys.\  B {\bf 695} (2004) 313
  [arXiv:hep-ph/0402225].
  %%CITATION = NUPHA,B695,31
G.~Moreau and J.~I.~Silva-Marcos,
  %``Flavour physics of the RS model with KK masses reachable at LHC,''
  JHEP {\bf 0603} (2006) 090
  [arXiv:hep-ph/0602155].
  %%CITATION = JHEPA,0603,090;%%
%\cite{Agashe:2006iy}

\bibitem{preNMFV}
  K.~Agashe, G.~Perez and A.~Soni,
  %``Flavor structure of warped extra dimension models,''
  Phys.\ Rev.\  D {\bf 71} (2005) 016002
  [arXiv:hep-ph/0408134]; 
  %%CITATION = PHRVA,D71,016002;%%
  K.~Agashe, A.~E.~Blechman and F.~Petriello,
  %``Probing the Randall-Sundrum geometric origin of flavor with lepton  flavor
  %violation,''
  Phys.\ Rev.\  D {\bf 74} (2006) 053011
  [arXiv:hep-ph/0606021].
  %%CITATION = PHRVA,D74,053011;%%

\bibitem{Contino:2006nn}
  R.~Contino, T.~Kramer, M.~Son and R.~Sundrum,
  %``Warped/Composite Phenomenology Simplified,''
  JHEP {\bf 0705} (2007) 074
  [arXiv:hep-ph/0612180].
  %%CITATION = JHEPA,0705,074;%%

  
\bibitem{Cacciapaglia:2007fw}
  G.~Cacciapaglia, C.~Csaki, J.~Galloway, G.~Marandella, J.~Terning and A.~Weiler,
  %``A GIM Mechanism from Extra Dimensions,''
  arXiv:0709.1714 [hep-ph].
  %%CITATION = ARXIV:0709.1714;%%


\bibitem{Fitzpatrick:2007sa}
  A.~L.~Fitzpatrick, G.~Perez and L.~Randall,
  %``Flavor from Minimal Flavor Violation & a Viable Randall-Sundrum Model,''
  arXiv:0710.1869 [hep-ph].
  %%CITATION = ARXIV:0710.1869;%%


%\cite{Nelson:2000sn}
\bibitem{Nelson:2000sn}
  A.~E.~Nelson and M.~J.~Strassler,
  %``Suppressing flavor anarchy,''
  JHEP {\bf 0009} (2000) 030
  [arXiv:hep-ph/0006251].
  %%CITATION = JHEPA,0009,030;%%



%\cite{Froggatt:1978nt}
\bibitem{Froggatt:1978nt}
  C.~D.~Froggatt and H.~B.~Nielsen,
  %``Hierarchy Of Quark Masses, Cabibbo Angles And CP Violation,''
  Nucl.\ Phys.\  B {\bf 147} (1979) 277.
  %%CITATION = NUPHA,B147,277;%%


%\cite{Leurer:1993gy}
\bibitem{Leurer:1993gy}
  M.~Leurer, Y.~Nir and N.~Seiberg,
  %``Mass matrix models: The Sequel,''
  Nucl.\ Phys.\  B {\bf 420} (1994) 468
  [arXiv:hep-ph/9310320].
  %%CITATION = NUPHA,B420,468;%%

%\cite{Ramond:1993kv}
\bibitem{Ramond:1993kv}
  P.~Ramond, R.~G.~Roberts and G.~G.~Ross,
  %``Stitching the Yukawa quilt,''
  Nucl.\ Phys.\  B {\bf 406} (1993) 19
  [arXiv:hep-ph/9303320].
  %%CITATION = NUPHA,B406,19;%%



%\cite{Leurer:1992wg}
\bibitem{Leurer:1992wg}
  M.~Leurer, Y.~Nir and N.~Seiberg,
  %``Mass matrix models,''
  Nucl.\ Phys.\  B {\bf 398} (1993) 319
  [arXiv:hep-ph/9212278].
  %%CITATION = NUPHA,B398,319;%%


%\cite{Buras:2001ra}
\bibitem{Buras:2001ra}
  A.~J.~Buras, S.~Jager and J.~Urban,
  %``Master formulae for Delta(F) = 2 NLO-QCD factors in the standard model  and
  %beyond,''
  Nucl.\ Phys.\  B {\bf 605}, 600 (2001)
  [arXiv:hep-ph/0102316].
  %%CITATION = NUPHA,B605,600;%%%
 

\bibitem{RGE}
  M.~Ciuchini, E.~Franco, V.~Lubicz, G.~Martinelli, 
  I.~Scimemi and L.~Silvestrini,
  %``Next-to-leading order QCD corrections to Delta(F) = 2 effective
  %Hamiltonians,''
  Nucl.\ Phys.\  B {\bf 523}, 501 (1998)
  [arXiv:hep-ph/9711402],
  %%CITATION = NUPHA,B523,501;%%%\cite{Buras:2000if}
%\bibitem{Buras:2000if}
  A.~J.~Buras, M.~Misiak and J.~Urban,
  % Two-loop QCD anomalous dimensions of flavour-changing 
  % four-quark  operators
  % within and beyond the standard model,''
  Nucl.\ Phys.\  B {\bf 586}, 397 (2000)
  [arXiv:hep-ph/0005183],
  %%CITATION = NUPHA,B586,397;%%
   

  %\cite{Becirevic:2001xt}
\bibitem{lattice}
  D.~Becirevic, V.~Gimenez, G.~Martinelli, M.~Papinutto and J.~Reyes,
  %``B-parameters of the complete set of matrix elements of Delta(B) = 2
  %operators from the lattice,''
  JHEP {\bf 0204}, 025 (2002)
  [arXiv:hep-lat/0110091].
  %%CITATION = JHEPA,0204,025;%%

\bibitem{Misiak:2006zs}
  M.~Misiak {\it et al.},
  %``The first estimate of B(anti-B --> X/s gamma) at O(alpha(s)**2),''
  Phys.\ Rev.\ Lett.\  {\bf 98}, 022002 (2007)
  [arXiv:hep-ph/0609232].
  %%CITATION = PRLTA,98,022002;%%

\bibitem{Misiak}
  M.~Misiak and M.~Steinhauser,
  %``NNLO QCD corrections to the anti-B --> X/s gamma matrix elements using
  %interpolation in m(c),''
  Nucl.\ Phys.\  B {\bf 764} (2007) 62
  [arXiv:hep-ph/0609241];
  %%CITATION = NUPHA,B764,62;%%
  M.~Misiak and M.~Steinhauser, private communication.

\bibitem{Colangelo:1999kr}
  G.~Colangelo, G.~Isidori and J.~Portoles,
  %``Supersymmetric contributions to direct CP violation in K --> pi pi  gamma
  %decays,''
  Phys.\ Lett.\  B {\bf 470}, 134 (1999)
  [arXiv:hep-ph/9908415].

\bibitem{Buras}
%%CITATION = PHLTA,B470,134;%%
  A.~J.~Buras, G.~Colangelo, G.~Isidori, A.~Romanino and L.~Silvestrini,
  %``Connections between epsilon'/epsilon and rare kaon decays in
  %supersymmetry,''
  Nucl.\ Phys.\  B {\bf 566}, 3 (2000)
  [arXiv:hep-ph/9908371],
  %%CITATION = NUPHA,B566,3;%%%
  G.~D'Ambrosio, G.~Isidori and G.~Martinelli,
  %``Direct CP violation in K --> 3pi decays induced by SUSY chromomagnetic
  %penguins,''
  Phys.\ Lett.\  B {\bf 480}, 164 (2000)
  [arXiv:hep-ph/9911522].
  %%CITATION = PHLTA,B480,164;%%


\bibitem{LFVops}
W.~Buchmuller and D.~Wyler,
  %``CP Violation And R Invariance In Supersymmetric Models Of Strong And
  %Electroweak Interactions,''
  Phys.\ Lett.\  B {\bf 121} (1983) 321.
  %%CITATION = PHLTA,B121,321;%% 
  R.~Kitano, M.~Koike and Y.~Okada,
  %``Detailed calculation of lepton flavor violating muon electron  conversion
  %rate for various nuclei,''
  Phys.\ Rev.\  D {\bf 66}, 096002 (2002)
  [Erratum-ibid.\  D {\bf 76}, 059902 (2007)]
  [arXiv:hep-ph/0203110]. 
  %%CITATION = PHRVA,D66,096002;%%

%\cite{Hisano:1995cp}
\bibitem{Hisano:1995cp}
  J.~Hisano, T.~Moroi, K.~Tobe and M.~Yamaguchi,
  %``Lepton-Flavor Violation via Right-Handed Neutrino Yukawa Couplings in
  %Supersymmetric Standard Model,''
  Phys.\ Rev.\  D {\bf 53} (1996) 2442
  [arXiv:hep-ph/9510309].
  %%CITATION = PHRVA,D53,2442;%%


%\cite{Willmann:1998gd}
\bibitem{Willmann:1998gd}
  L.~Willmann {\it et al.},
  %``New bounds from searching for muonium to antimuonium conversion,''
  Phys.\ Rev.\ Lett.\  {\bf 82} (1999) 49
  [arXiv:hep-ex/9807011].
  %%CITATION = PRLTA,82,49;%%


\bibitem{PDG}
  W.~M.~Yao {\it et al.}  [Particle Data Group],
  %``Review of particle physics,''
 J.\ Phys.\ G {\bf 33} (2006) 1.
 %%CITATION = JPHGB,G33,1;%%


%\cite{Hayasaka:2007vc}
\bibitem{Hayasaka:2007vc}
  K.~Hayasaka {\it et al.}  [Belle Collaboration],
  %``New search for tau --> mu gamma and tau --> e gamma decays at Belle,''
  arXiv:0705.0650 [hep-ex].
  %%CITATION = ARXIV:0705.0650;%%

%\cite{Aubert:2005wa}
\bibitem{Aubert:2005wa}
  B.~Aubert {\it et al.}  [BABAR Collaboration],
  %``Search for lepton flavor violation in the decay $\tau^\pm \to e^\pm
  %\gamma$,''
  Phys.\ Rev.\ Lett.\  {\bf 96} (2006) 041801
  [arXiv:hep-ex/0508012].
  %%CITATION = PRLTA,96,041801;%%


%\cite{Bennett:2006fi}
\bibitem{Bennett:2006fi}
  G.~W.~Bennett {\it et al.}  [Muon G-2 Collaboration],
  %``Final report of the muon E821 anomalous magnetic moment measurement at
  %BNL,''
  Phys.\ Rev.\  D {\bf 73} (2006) 072003
  [arXiv:hep-ex/0602035].
  %%CITATION = PHRVA,D73,072003;%%

}
\end{thebibliography}
\end{document}